\documentclass[twocolumn,showpacs,preprintnumbers,amsmath,amssymb,floatfix]{revtex4-1}

\usepackage{color}
\usepackage{graphicx}
\usepackage{dcolumn}
\usepackage{bm}

\begin{document}

\title{ Extended Glauber Model of Antiproton-Nucleus Annihilation
  \\ for All Energies and Mass Numbers}

\author{Teck-Ghee Lee$^1$ and Cheuk-Yin Wong$^2$}

\affiliation{$^1$Department of Physics, Auburn University, Auburn, AL
  36849, U.S.A.}

\affiliation{$^2$Physics Division, Oak Ridge National Laboratory, Oak
  Ridge, TN 37831, U.S.A.}

\def\bb    #1{\hbox{\boldmath${#1}$}}

\begin{abstract}
Previous analytical formulas in the Glauber model for high-energy
nucleus-nucleus collisions developed by Wong are utilized and extended
to study antiproton-nucleus annihilations for both high and low
energies, after taking into account the effects of Coulomb and nuclear
interactions, and the change of the antiproton momentum inside a
nucleus.  The extended analytical formulas capture the main features
of the experimental antiproton-nucleus annihilation cross sections for
all energies and mass numbers.  At high antiproton energies, they
exhibit the granular property for the lightest nuclei and the
black-disk limit for the heavy nuclei.  At low antiproton energies,
they display the effect of antiproton momentum increase due to the
nuclear interaction for light nuclei, and the effect of
magnification due to the attractive Coulomb interaction for heavy
nuclei.
\end{abstract}

\maketitle

\section{Introduction}
Recently, there has been interest in the interaction of antimatter with
matter, as it is central to our understanding of the basic structure
of matter and the matter-antimatter asymmetry in the Universe.  On 
the one hand, the land-based Facility forAntiprotons and Ions Research 
(FAIR) at Darmstadt\cite{Fai09,Pan13} and the Antiproton Decelerator 
(AD) at CERN \cite{Mau99} have been designed to
probe the interaction of antiprotons with matter at various energies
and environments.  The orbiting Payload for Antimatter Matter 
Exploration and Light-Nuclei Astrophysics (PAMELA)
\cite{PAMELA} and the Alpha Magnetic Spectrometer (AMS) \cite{AMS02}
measure the intensity of antimatter in outer space.  They have
provided interesting hints on the presence of extra-terrestrial
antimatter sources in the Universe.  In support of these facilities
for antimatter investigations, it is of interest to examine here the
antiproton-nucleus annihilation cross sections that represent an
important aspect of the interaction between antimatter and matter.

To date, significant experimental and theoretical efforts have been
put forth to understand the process of annihilation between $\bar{p}$
and various nuclei across the periodic table from low energy to high
energies.  On the experimental side, annihilation cross sections for
${\bar p} A$ collisions, $\sigma_{\rm ann}^{\bar p A}$, have been
measured at the Low-Energy Antiproton Ring (LEAR) at CERN
\cite{Bia11}-\cite{Al} and compiled in Ref.\ \cite{Bia11}, and at the
Detector of Annihilations (DOA) at Dubna \cite{Kuz94}. A surprising
difference in the behavior at high and low energies has been detected.
In light nuclei (H, $^2$H and $^4$He), the $\bar{p}A$ annihilation cross
sections at $\bar{p}$ momenta below 60 MeV/$c$ have comparable values,
whereas at momenta greater than 500 MeV/$c$ the $\bar{p}$-nucleus
annihilation cross sections increases approximately linearly with the
mass number $A$ for the lightest nuclei \cite{Bia11, Ber96,
  Zen99a,Zen99b}.  On the other hand, for collisions with heavy nuclei
at low energies, the annihilation cross section exhibits large
enhancements that are much greater than what one would expect just
from the geometrical radii alone \cite{Bia11}.
 
On the theoretical side, a theoretical optical potential based on the
Glauber model \cite{Gla59,Gla70} has been developed by Kuzichev,
Lepikhin and Smirnitsky to investigate the antiproton annihilation
cross sections on Be, C, Al, Fe, Cu, Cd, and Pb nuclei in the momentum
range 0.70-2.50 GeV/$c$ \cite{Kuz94}.  In this range of relatively
high antiproton momenta, the Glauber model gives a good agreement with
the experimental data, with the exception of the deviations at the
momentum of 0.7 GeV/$c$ for heavy nuclei.  Their study suggested that
the $A$ dependence of the annihilation cross sections is influenced by
Coulomb interaction at low momenta.  In another analysis, Batty,
Friedman, and Gal have developed a unified optical potential approach
for low-energy $\bar{p}$ interactions with protons and with various
nuclei \cite{Gal00,Bat01}. Starting with a simple optical potential
$\bar v$ determined by comparison with $\bar{p}$-$p$ experimental
data, a density-folded optical potential $V_{\rm opt}=\rho \bar v$ was
formulated for the collision of the $\bar{p}$-nucleus system.  They
found that even though the density-folding potential reproduces
satisfactorily the $\bar{p}$ atomic level shifts and widths across the
periodic table for $A$$>$10 and the few annihilation cross sections
measured on Ne, it does not work well for He and Li. They attributed
this discrepancy to the spin and the isospin averaging and the
approximations made in the constructions of the optical potentials. An
extended black-disk strong-absorption model has also been considered
to account for the Coulomb focusing effect for low-energy $\bar{p}$
interactions with nuclei and a fair agreement with the measured
annihilation cross sections was achieved \cite{Bat01}.

There are many puzzling features of the $\bar{p}A$ annihilation cross
sections that are not yet well understood.  With respect to the mass
dependence, why at high antiproton incident momenta do the cross
sections increase almost linearly with the mass number $A$ for the
lightest nuclei, but with approximately $A^{2/3}$ as the mass number
increases?  Why in the low antiproton momentum region do the
annihilation cross sections not rise with $A$ as anticipated but are
comparable for $p$, $^2$H, and He, and become subsequently greatly
enhanced as $A$ increases further into the heavy nuclei region? With
respect to the energy dependence, how does one understand the energy
dependence of the $\bar p A$ annihilation cross sections and the
relationship between the energy dependence of the $\bar p p$ cross
sections and the $\bar p A$ cross sections?  In what roles do the
residual nuclear interaction and the long-range Coulomb interaction
interplay in the cross sections as a function of charge numbers and
antiproton momenta?  We would like to design a model in which these
puzzles can be brought up for a close examination.

In order to be able to describe the collision with all nuclei,
including deuteron, it is clear that the model needs to be
microscopic, with the target nucleon number $A$ appearing as an
important discrete degree of freedom.  Furthermore, with the
conservation of the baryon number, an antiproton projectile can only
annihilate with a single target nucleon.  The annihilation process
occurs between the projectile antiproton and a target nucleon locally
within a short transverse range along the antiproton trajectory.  This
process of annihilation occurring in a short range along the
antiproton trajectory is similar in character to the high-energy $pA$
reaction process in which the incident project $p$ interacts with
target nucleons along its trajectory.  In the case of high energy
collisions, the trajectory of the incident projectile can be assumed
to be along a straight line, and the reaction process can be properly
described by the Glauber multiple collision model
\cite{Gla59,Gla70,Won84,Won94,Kuz94}.

Previously, analytical formulas for high-energy nucleon-nucleus and
nucleus-nucleus collisions in the Glauber multiple collision model
have been developed by Wong for the reaction cross section in $pA$
collisions as a function of the basic nucleon-nucleon cross section
\cite{Won84,Won94}. The analytical formula involves a discrete sum of
probabilities whose number of terms depend on the number of target
nucleons as a discrete degree of freedom.  They give the result that
the $pA$ reaction cross section is proportional to $A$ for small $A,$
and approaches the black-disk limit of $A^{2/3}$ for large $A$,
similar to the mass-dependent feature of the $\bar p A$ annihilation
cross sections at high energies mentioned earlier.  Hence, it is
reasonable to utilize these analytical formulas and concepts in the
Glauber multiple collision model \cite{Won94} to provide a description
of the annihilation process in $\bar p A$ reactions.

As the analytical results in the Glauber model \cite{Won84,Won94}
pertain to high-energy nucleon-nucleus collisions with a straight-line
trajectory, the model must be extended and amended to make them
applicable to low-energy $\bar p$-nucleus annihilations.  The incident
antiproton is subject to the initial-state Coulomb interaction
\cite{Kuz94,Bat01}.  The antiproton trajectory deviates from a
straight line in low-energy collisions.  Before the antiproton comes
into contact with the nucleus, the antiproton trajectory is pulled
towards the target nucleus, resulting in a magnifying lens effect (or
alternatively a Coulomb focusing effect \cite{Kuz94,Bat01}) that
enhances greatly the annihilation cross section.  Furthermore, the
antiproton is subject to the nuclear interaction that changes the
antiproton momentum in the interior of the nucleus.  The change of
antiproton momentum is especially important in low energy
annihilations of light nuclei because of the strong momentum
dependence of the basic $\bar p p$ annihilation cross section.  It is
necessary to modify the analytical formulas to take into account these
effects so that they can be applied to $\bar p$-nucleus annihilations
for all energies and mass numbers.  Success in constructing such an
extended model will allow us to resolve the puzzles we have just
mentioned.

This paper is organized as follows.  In Sec. II, we review and
summarize previous results in the Glauber model for high-energy
nucleon-nucleus collisions, to pave the way for its application to
$\bar p$-nucleus collisions.  Analytical formulas are written out for
the $\bar p$-nucleus annihilation cross sections in terms of basic
$\bar p $-nucleon annihilation cross section, $\sigma_{\rm ann}^{\bar
  p-{\rm nucleon}}$.  We use the quark model to relate $\sigma_{\rm
  ann}^{\bar p n}$ to $\sigma_{\rm ann}^{\bar p p}$ so that it suffices
to use only $\sigma_{\rm ann}^{\bar p p}$ to evaluate the $\bar
p$-nucleus cross section.  In Sec. III, we represent the basic
$\bar p p$ annihilation cross section by a $1/v$ law.  In Sec. IV,
we extend the Glauber model to study antiproton-nucleus annihilations
at both high and low energies, after taking into account the effects
of Coulomb and nuclear interactions, and the change of the antiproton
momentum inside a nucleus.  In Sec. V, we compare the results of
the analytical formulas in the extended Glauber model to experimental
data.  We find that these analytical formulas capture the main
features of the experimental antiproton-nucleus annihilation cross
sections for all energies and mass numbers.  Finally, we conclude the
present study with some discussions in Sec. VI.

\section{Glauber Model for $\bar p$-nucleus Annihilation at High Energies}

We shall first briefly review and summarize the analytical formulas in
the Glauber multiple collision model \cite{Won84,Won94} for its
application to $\bar p$-nucleus annihilations at high energies.  The
Glauber model assumes that the incident antiproton travels along a
straight line at a high energy and makes multiple collisions with
target nucleons along its way.  The target nucleus can be represented
by a density distribution.  The integral of the density distribution
along the antiproton trajectory gives the thickness function which, in
conjunction with the basic antiproton-nucleon annihilation cross
section $\sigma_{\rm ann}^{\bar p p}$, determines the probability for
an antiproton-nucleon annihilation and consequently the $\bar
p$-nucleus annihilation cross section \cite{Gla59,Gla70,Won84,Won94}.
  
To be specific, we consider a target nucleus $A$ with a thickness
function $T_A(\bb b_A)$ and mass number $A$, and a projectile
antiproton with a thickness function $T_B(\bb b_B)$ and a mass number
$B$=1.  The integral of all thickness functions are normalized to
unity.  For simplicity, we shall initially not distinguish between the
annihilation of a proton or a neutron.  Refinement to allow for
different annihilation cross sections will be generalized at the end
of this section.

According to Eq. (12.8) of \cite{Won94}, in general, the thickness
function $T(\bb b)$ for the annihilation between the projectile
antiproton $B$ and a nucleon in the target nucleus $A$ at high
energies along a straight-line trajectory at the transverse coordinate
$\bb b$ is given by
\begin{eqnarray}
T(\bb b) =\!\! \int\!\! \bb db_A \!\!\int\!\! \bb db_B T_A(\bb b_A) T_B(\bb b_B) t_{\rm ann}(\bb b - \bb b_A \bb+\bb b_B),~~
\label{1}
\end{eqnarray}
where $t_{\rm ann}(\bb b - \bb b_A \bb+\bb b_B )$ is the annihilation
thickness function, specifying the probability distribution at the
relative transverse coordinate $\bb b - \bb b_A \bb+\bb b_B$ for the
annihilation of a target nucleon at $\bb b_A$ with an antiproton at $\bb b_B$.

The thickness function $t_{\rm ann}(\bb b)$ for $\bar p p$
annihilation at $\bb b$ can be represented by a Gaussian with a
standard deviation $\beta_{\bar p p}$,
\begin{eqnarray}
t_{\rm ann}(\bb b)=\frac{1}{2\pi \beta_{\bar p p}^2}\exp \{ - \frac{\bb b^2}{2 \beta_{\bar p p}^2} \}.
\end{eqnarray}
The cross section for a $ \bar p p$ annihilation is then given by
\begin{eqnarray}
\sigma_{\rm ann}^{ \bar p p} &=& \int d\bb b \,(\pi {\bb b}^2)\, t_{\rm ann}(\bb b)
=2\pi  \beta_{\bar p p}^2 ,
\end{eqnarray}
where $d\bb b=2\pi b db$.  Therefore, the standard deviation
$\beta_{\bar p p}$ in the $\bar p p$ annihilation thickness function
is related to the $\bar p p$ annihilation cross section by
\begin{eqnarray}
\beta_{\bar p p}^2 =\frac{\sigma_{\rm ann}^{ \bar p p} }{2\pi}.
\end{eqnarray}
In a $\bar p$-$A $ collision at high energies, the probability for the
occurrence of an annihilation is $T(b) \sigma_{\rm ann}^{\bar pp}$.
The probability for no annihilation is $[1-T(b) \sigma_{\rm ann}^{\bar
    pp}]$.  With $A$ target nucleons, the annihilation cross section
in a $\bar p A$ collision at high energies, as a function of
$\sigma_{\rm ann}^{\bar pp}$, is therefore
\begin{eqnarray}
\sigma_{\rm ann}^{\bar pA}(\sigma_{\rm ann}^{\bar
    pp})=\int d\bb b \biggl  \{ 1 - [1-T(\bb b) \sigma_{\rm ann}^{\bar p p}]^{A} \biggr \} .
\label{5}
\end{eqnarray}

It should be noted that $\sigma_{\rm ann}^{\bar p p}$ depends on the
magnitude of the antiproton momentum relative to the target nucleons.
For example, in our later applications to extend the Glauber model to
low energies, the antiproton momentum at the moment of $\bar
p$-nucleon annihilation may be significantly different from the
incident antiproton momentum, and it becomes necessary to specify the
momentum dependence $\sigma_{\rm ann}^{\bar p p}$ in Eq.\ (\ref{5})
explicitly.  For brevity of notation, we shall not write out the
momentum dependence explicitly except when it is needed to avoid
momentum ambiguities.

Analytical expressions of $\sigma_{\rm ann}^{\bar pA}(\sigma_{\rm
  ann}^{\bar pp})$ can be obtained for simple thickness functions
\cite{Won84,Won94}.  If the thickness functions of $T_A$ and $T_B$ are
Gaussian functions with standard deviation $\beta_A$ and $\beta_B$
respectively, then Eq.\ (\ref{1}) gives
\begin{eqnarray}
T(\bb b)&=& \frac{1}{2\pi \beta^2}\exp \{ - \frac{ b^2}{2 \beta^2} \},
\label{6}
\end{eqnarray}
where
\begin{eqnarray}
\beta^2&=&\beta_A^2+\beta_B^2+\beta_{\bar p p}^2.
\label{8}
\end{eqnarray}
For this case with Gaussian thickness functions, Eq.\ (\ref{5}) then
gives the simple analytical formula  \cite{Won84,Won94}
\begin{eqnarray}
\sigma_{\rm ann}^{\bar pA}(\sigma_{\rm ann}^{\bar p p})
&=&2\pi \beta^2 \sum_{n=1}^{A} \
\frac{1-(1-f)^n}{n},
\label{9}
\end{eqnarray}
where
\begin{eqnarray}
f=\frac{\sigma_{\rm ann}^{\bar p p}}{2\pi \beta^2}
= \frac{\sigma_{\rm ann}^{ \bar p p }}{2\pi [\beta_A^2+\beta_B^2]+\sigma_{\rm ann}^{ \bar p p }}.
\label{10}
\end{eqnarray}
To check our theory, we apply the results first to the case of $A=1$, and we obtain
 \begin{eqnarray}
\sigma_{\rm ann}^{\bar p p}&=&2\pi \beta^2 
\left(\frac{1-(1-f)^1}{1}\right)
=\sigma_{\rm ann}^{\bar p p} ,
\end{eqnarray}
as it should be.  Next, for $\bar p^2$H collisions where $A=2$, we have
 \begin{eqnarray}
\sigma_{\rm ann}^{\bar p ^2H}
&=&2\pi \beta^2 
\left[2 f-\frac{f^2}{2}\right].
\end{eqnarray}
The situation depends on the size of $\beta_{\bar p p}$ (or
$\sigma_{\rm ann}^{\bar p p}$), relative to $\beta_A$ and $\beta_B$.
There are two different limits of $\sigma_{\rm ann}^{\bar p p}$ in
comparison with $2\pi [\beta_A^2+\beta_B^2]$.  If $\sigma_{\rm
  ann}^{\bar p p} \ll 2\pi [\beta_A^2+\beta_B^2] $ then $f \to 0 $ and
\begin{eqnarray}
\sigma_{\rm ann}^{\bar p ^2H}
\sim   2 \sigma_{\rm ann}^{\bar p p},
\end{eqnarray}
which exhibits the granular property of the nucleus when the basic
antiproton-nucleon cross section is much smaller than the radius of the
nucleus.  On the other hand, if $ \sigma_{\rm ann}^{\bar p p} \gg 2\pi
[\beta_A^2+\beta_B^2] $, then $f \to 1$ and the cross section become
 \begin{eqnarray}
\sigma_{\rm ann}^{\bar p ^2H}\sim
\frac{3}{2}\sigma_{\rm ann}^{\bar p p}.
\label{14}
\end{eqnarray}
In actual comparison with experimental data, we use $\beta_B$ = 0.68
fm, and we parametrize $\beta_A$ = $r'_0 A^{1/3}/\sqrt{3}$.  The
standard root-mean-squared radius parameter $r'_0$ is of order 1 fm
(see Table I below).  The $\bar p p$ annihilation cross section
$\sigma_{\rm ann}^{\bar p p}$ is about 50 mb at $p_{\rm lab}^{\bar
  p}$=2 GeV/$c$ and about 1000 mb at $p_{\rm lab}^{\bar p }$=50 MeV/$c$.
Thus, $f \ll 1$ for $p_{\rm lab}^{\bar p p}$=2 GeV/$c$ and $f \sim 1$
for $p_{\rm lab}^{\bar p}$=50 MeV/$c$.  If the Glauber model remains
valid for the whole momentum range, then one expects that
\begin{eqnarray}
{\sigma_{\rm ann}^{\bar p ^2H}}/
{ \sigma_{\rm ann}^{\bar p p}}\bigg|_{\rm Glauber ~model}\!\!\!\!\!\! \sim 
\begin{cases} 
2   &   \!\! {\rm for~ high~}{\bar p}{\rm~momenta},     \cr
3/2 &  \!\!{\rm for~ low~~ } {\bar p}{\rm~momenta}.~~~~  
\label{15}
\end{cases}
\end{eqnarray}
The experimental data indicate
\begin{eqnarray}
{\sigma_{\rm ann}^{\bar p ^2H}}/
{ \sigma_{\rm ann}^{\bar p p}}\bigg|_{\rm experimental}  \!\!\!\!\sim 
\begin{cases} 
2   &    \!\!  {\rm for~ high~}{\bar p}{\rm~momenta},     \cr
1 &   \!\! {\rm for~ low~~ } {\bar p}{\rm~momenta}.~~~~  \cr
\end{cases}
\end{eqnarray}

The predicted ratio of ${\sigma_{\rm ann}^{\bar p ^2H}}/ { \sigma_{\rm
    ann}^{\bar p p}}$ appears correct for the high-energy region.
However, for low-momentum $\bar p$ annihilations, we shall see that
there are important modifications that must be made to extend the
Glauber model to the low-momentum region, and these modifications will
alter the $\sigma_{\rm ann}^{\bar p ^2H}/ \sigma_{\rm ann}^{\bar p p}$
ratio in that region.

As the nuclear mass number $A$ increases, the density distribution of
the nucleus become uniform.  The thickness function for the collision
of $\bar p$ with a heavy nucleus can be approximated by using a
sharp-cut-off distribution of the form (see Ref. \cite{Won84,Won94})
\begin{eqnarray}
T(b) = \frac{3\sqrt{(R_c^2-b^2)}}{2\pi R_c^3} \theta(R_c-b),
\label{17}
\end{eqnarray}
where the contact radius can be taken to be
\begin{eqnarray}
R_c=R_A+R_B+R_{p\bar p}.
\label{12}
\end{eqnarray}
With this sharp-cut-off distribution, Eq.\ (\ref{5}) leads to the
cross section given by \cite{Won84,Won94}
\begin{eqnarray}
 \sigma_{\rm ann}^{\bar p A}&&(\sigma_{\rm ann}^{\bar p p})
= \pi R_c^2 
\label{19}\\
 &&\times \biggl ( 1 +\frac{2}{F^2}\left[ \frac{1-(1-F)^{A+2}}{A+2}-\frac{1-(1-F)^{A+1}}{A+1}\right]\biggr ),\nonumber
\end{eqnarray} 
where $F$ is a dimensionless ratio,
\begin{eqnarray}
F=\frac{\sigma_{\rm ann}^{\bar p p}}{2\pi R_c^2/3}.
\label{20a}
\end{eqnarray}
The $\bar p p$ annihilation radius $R_{p\bar p}$ in Eq.\ (\ref{12})
can be calibrated to be $R_{p\bar p} = \sqrt{(3 \sigma_{\rm ann}^{\bar
    p p} / 2 \pi)}$ by using the above equation (\ref{19}) for the
case of $\bar p p$ collision as point nucleons. 

In the foregoing discussions, we assume that $\sigma_{\rm ann}^{\bar p
  p}$ and $\sigma_{\rm ann}^{\bar p n}$ are the same.  They actually
differ by about 20\%.  We can take into account different annihilation
cross sections and thickness functions $T_{\bar p p}$ and $T_{\bar p
  n}$ for protons and neutrons.  Equation (\ref{5}) can be generalized
to become
\begin{eqnarray}
 \sigma_{\rm ann}^{\bar p A}(\sigma_{\rm ann}^{\bar p-{\rm nucleon}})
&=& \int d\bb b \biggl  
\{ 1 - [1-T_{\bar p p}(\bb b) \sigma_{\rm ann}^{\bar p p}]^{Z}
\nonumber \\
&&\times [1-T_{\bar p n}(\bb b) \sigma_{\rm ann}^{\bar p n}]^{N} \biggr \}
\nonumber \\
& &  \!\!\!\! \!\!\!\! \!\!\!\! \!\!\!\! \!\!\!\! \!\!\!\!=
\sum_{i=0}^Z \sideset{}{^\prime}\sum_{j = 0}^N
\left(\frac{(-1)^{1+i+j}Z!N!}{(Z-i)!(N-j)! i!j!}\right)  
\nonumber \\
&&  \!\!\!\! \!\!\!\! \!\!\!\! \!\!\!\! \!\!\!\!  \!\!\!\!\!\!\!\!\times
 (\sigma_{\rm ann}^{\bar p  p})^i 
(\sigma_{\rm ann}^{\bar p  n})^j
\int d \bb b [T_{\bar p p}(\bb b)]^i[T_{\bar p n}(\bb b)]^j,
\label{100}
\end{eqnarray} 
where the argument $ \sigma_{\rm ann}^{\bar p -{\rm nucleon}}$ on the
left-hand side stands for $\sigma_{\rm ann}^{\bar p p}$ and $
\sigma_{\rm ann}^{\bar p n}$, and the summation $\sum'_j$ allows for
all cases except when $i=j=0$.

For small and moderate mass numbers $A <$ 40, the thickness functions
$T_{\bar p x} (b)$ can be assumed to be a Gaussian function with a standard
deviation $\beta_x$, and Eq.\ (\ref{100}) becomes
\begin{eqnarray}
 \sigma_{\rm ann}^{\bar p A}(\sigma_{\rm ann}^{\bar p-{\rm nucleon} })
&=& 2\pi \sum_{i=0}^Z \sideset{}{^\prime}\sum_{j = 0}^N
\left ( \frac{(-1)^{1+i+j}Z!N!}{(Z-i)!(N-j)! i!j!}\right ) 
\nonumber \\
&& \!\!\!\!\!\!\!\!\times
 \left(\frac{\sigma_{\rm ann}^{\bar p  p}}{2 \pi \beta_p^2}\right)^i 
\left(\frac{\sigma_{\rm ann}^{\bar p  n}}{2 \pi \beta_n^2}\right)^j
\left(\frac{\beta_p^2\beta_n^2}{i\beta_n^2+j\beta_p^2}\right), ~~ ~~ 
\label{21a}
\end{eqnarray}   
where $\beta_p^2 = \beta_A^2+ \beta_B^2 + \sigma_{\rm ann}^{\bar p p}/2\pi$ 
and $\beta_n^2 = \beta_A^2+ \beta_B^2 + \sigma_{\rm ann}^{\bar p n}/2\pi$.

On the other hand, as the value of $A$ increases, the $T_{\bar p
  x}(b)$ function becomes a uniform distribution. With this
sharp-cut-off distribution, Eq.\ (\ref{100}) leads to the cross
section given by
\begin{eqnarray}
 \sigma_{\rm ann}^{\bar p A}(\sigma_{\rm ann}^{\bar p -{\rm nucleon}})
&=& \pi R^2_{c,n}\sum_{i=0}^Z \sideset{}{^\prime}\sum_{j = 0}^N
\left(\frac{(-1)^{1+i+j}Z!N!}{(Z-i)!(N-j)! i!j!}\right)  
\nonumber \\
&& \!\!\! \times\left(\frac{3\sigma_{\rm ann}^{\bar p  p}}{2 \pi R_{c,p}^2}\right)^i 
\left(\frac{3\sigma_{\rm ann}^{\bar p  n}}{2 \pi R_{c,n}^2}\right)^j I(i,j,a),
~~ ~~~~
\label{22a}
\end{eqnarray} 
where $R_{c,x} = R_A+ R_B + \sqrt{(3\sigma_{\rm ann}^{\bar p x}/2\pi)}$
and $ x = \{p,n\}$. The function $I(i,j, a)$ is
\begin{eqnarray}
I(i,j, a)&=& \int_0^{R_{c,n}} \frac {2bdb}{R^2_{c,n}}
\left(1-\frac{b^2}{R_{c,p}^2}\right)^{i/2}
\left(1-\frac{b^2}{R_{c,n}^2}\right)^{j/2}
\nonumber\\
&=&
\int_0^{1} dy(1-ay)^{i/2}(1-y)^{j/2},
\end{eqnarray}
where $a = R^2_{c,n}/R^2_{c,p} < 1 $. The $I(i,j, a)$ function is
evaluated numerically.

\section{The Basic $\bar p p$ annihilation cross section 
$\sigma_{\rm ann}^{\bar p p}$}

In the previous section, analytical formulas have been written out for
the annihilation cross sections in the collision of an antiproton in
terms of $\sigma_{\rm ann}^{\bar p -{\rm nucleon}}$ (i) for a light
target nucleus in Eq.\ (\ref{21a}) [or Eq.\ (\ref{9}) if we assume
  $\sigma_{\rm ann}^{\bar p p} \sim \sigma_{\rm ann}^{\bar p n}$], and
(ii) for a heavy nucleus in Eq. (\ref{22a}) [or Eq. (\ref{19}) if we
  assume $\sigma_{\rm ann}^{\bar p p} \sim \sigma_{\rm ann}^{\bar p
    n}$].  The evaluation of the $\bar p A$ annihilation cross section
will require knowledge of the basic $\sigma_{\rm ann}^{\bar p-{\rm
    nucleon}}$ annihilation cross section.

From the quark model, a proton $p$ consists of $uud$, a neutron $n$
consists of $udd$, and a $\bar p$ is $\bar u \bar u \bar d$. If only
flavor and antiflavor can annihilate, one would expect the probability
of $\bar p$-$n$ annihilation equals (4/5) times the probability of
$\bar p$-$p$ annihilation, that is,
\begin{eqnarray}
\sigma_{\rm ann}^{\bar p n} = (4/5) \sigma_{\rm ann}^{\bar p p}.
\label{24a}
\end{eqnarray}
The experimental value for the ratio $(\sigma_{\rm ann}^{\bar p
  n})_{^2H}/(\sigma_{\rm ann}^{\bar p p})_{^2H}$ for the annihilation inside a
deuteron has been determined by Kalogeropoulos and Tzanakos \cite{D1}
to be $0.745~\pm~0.016$ and $0.863~\pm~0.016$ for $\bar p$ at rest and
in flight, respectively, giving an average of 0.804$\pm$0.022, in
approximate agreement with the prediction from the naive quark model.
We shall use the relation Eq.\ (\ref{24a}) in our subsequent data
analysis.  By relating $\sigma_{\rm ann}^{\bar p n}$ with $\sigma_{\rm
  ann}^{\bar p p}$ using Eq.\ (\ref{24a}), the evaluation of the $\bar
p A$ annihilation cross section will require only the basic $
\sigma_{\rm ann}^{\bar p p}$ annihilation cross section.

In our previous theoretical study in connection with the annihilation
lifetime of matter-antimatter molecules, we note that the
experimental data of the $\bar p p$ total cross section and the $\bar
p p$ elastic cross section as a function of the antiproton momentum for
a fixed proton target, $p_{\bar p{\rm lab}}$, are related by
\cite{Won11}
\begin{eqnarray}
\sigma_{\rm tot}=\sigma_{\rm elastic}+\frac{\sigma_0}{v},
\label{21}
\end{eqnarray}
where $v$ is the velocity of the antiproton 
\begin{eqnarray}
v=\frac{p_{{\bar p}{\rm lab}}}{\sqrt{p_{{\bar p} {\rm lab}}^2+m_{\bar p}^2}}.
\end{eqnarray}
 
As the difference of the experimental total cross section and the
elastic cross section, the second term in Eq.\ (\ref{21}) represents
the $\bar p p$ inelastic cross section.  It is essentially the
annihilation cross section as the latter dominates among the inelastic
channels.  The $\bar p p$ annihilation cross section can therefore be
parametrized in the form
\begin{eqnarray}
\sigma_{\rm ann}^{\bar p p}=\frac{ \sigma_0}{v}.
\label{17a}
\end{eqnarray}

 In the present work, we use the experimental $\bar p p$ annihilation cross
 sections directly to fine-tune the parameter $\sigma_0$.  We find
 that
\begin{eqnarray}
\sigma_0 = 43 {\rm ~mb}
\end{eqnarray}
gives a good description of the experimental $\sigma_{\rm ann}^{\bar p
  p}(p_{{\bar p} {\rm lab}})$ data as shown in Fig. 1.  For brevity of
notation, the quantity $p_{\bar p{\rm lab}}$ will be abbreviated as
$p_{\rm lab}$ in all figures.

Figure 1 indicates that the  $\bar p p$ annihilation cross
section has a strong momentum dependence.  It decreases by an order
of magnitude as the antiproton momentum increases from 30 to 600 MeV/$c$.

\begin{figure*}[h]
  \centering
\includegraphics[scale=0.65]{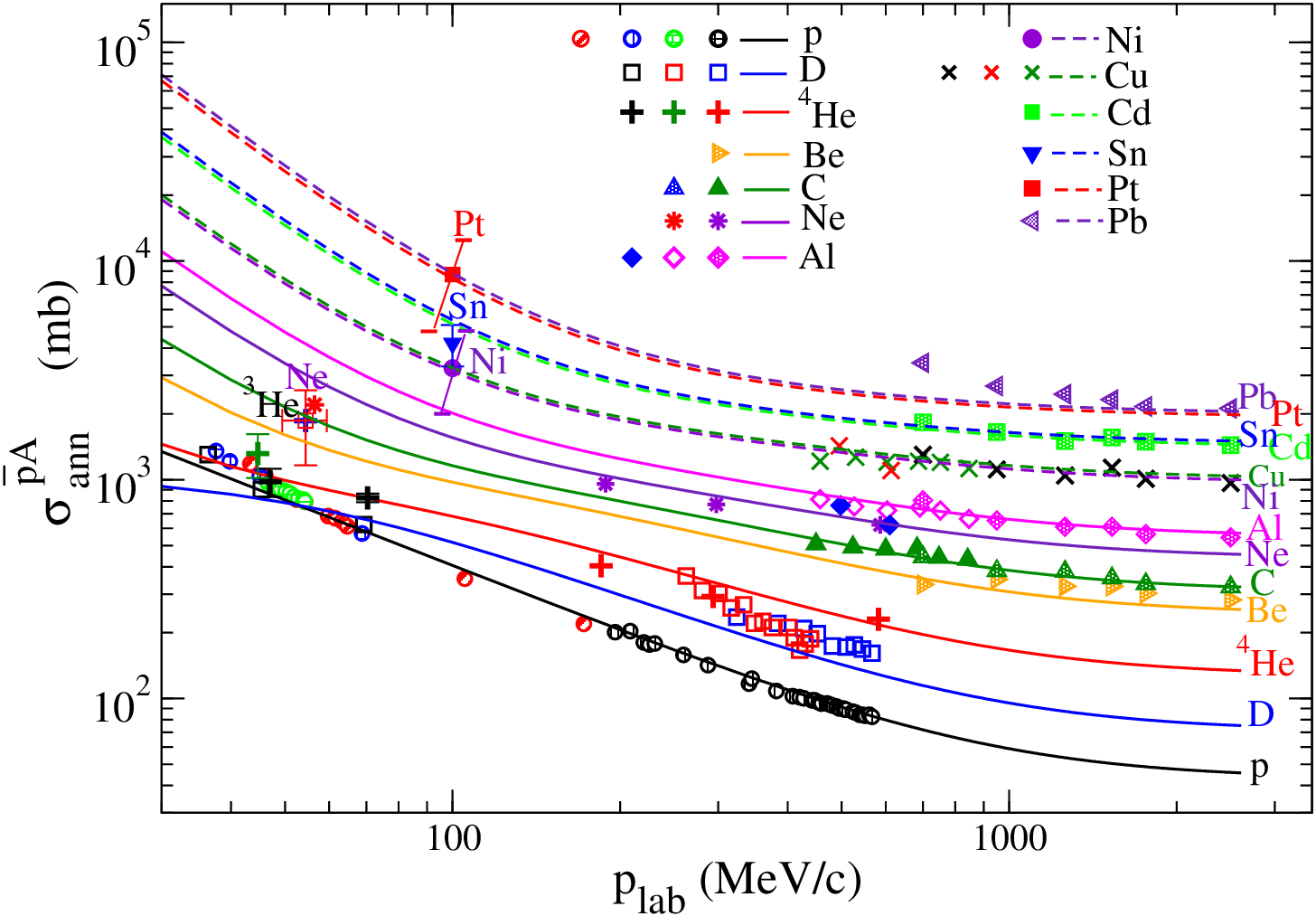}
\caption{(Color online) Antiproton-nucleus annihilation cross sections
  as a function of the antiproton momentum in the laboratory, $p_{\rm
    lab}\equiv p_{\bar p {\rm lab}}$, for different nuclear targets.
  The solid curve for the proton target nucleus is the $\sigma_0/v$
  phenomenological representation of the $\bar p p$ annihilation cross
  section in Eq.\ (\ref{17a}).  The other curves are the results from
  the extended Glauber model using the basic $\bar p p$ annihilation
  cross section as input data.  The solid curves are for Gaussian
  density distributions and the dashed curves are for uniform density
  distributions. The high-momentum data points are from \cite{Kuz94}.
  The other experimental data points are from the compilation of
  \cite{Bia11}, where the individual experimental sources can be
  found.  }
\end{figure*}

It should be pointed out that the $1/v$ law, Eq.\ (\ref{17a}), for the
inelastic (or annihilation) cross section of slow particles is well
known.  It was first obtained by Bethe \cite{Bet35} and is discussed
in text books \cite{Bla52,Lan58,Sat90} and other related work
\cite{Won97}.  It arises from multiplying the $S$-wave partial cross
section, $\pi/k^2$, by the transmission coefficient $T_0$ in passing
through an attractive potential well, and the transmission coefficient
$T_0$ is proportional to $k$ at low energies.  While the $1/v$
behavior is essentially an $S$-wave result, higher-$l$ partial waves
will gradually contribute as the antiproton momentum reaches the GeV/$c$
region.  It is nonetheless interesting to note that the simple $1/v$
law of Eq.\ (\ref{17a}) continues to provide a reasonable and efficient
description of the experimental data even in the GeV/$c$ region
\cite{Won11}.

Recently, an elaborate and model-independent coupled-channel
partial-wave calculation, solving the problem from first principle,
has been used to determine $\bar{p}$-$p$ scattering cross sections for
momenta below 0.925 GeV/$c$. The calculation yields excellent agreement
with the experimental $\bar{p}$-$p$ annihilation cross section for
energy ranging from 0.200 to 0.925 GeV/$c$ \cite{Zhou12}.  Refinement of
the present $\sigma_{\rm ann}^{\bar p p}$ description can be made, if
desired, but with additional complications.

\section{Extending the Glauber Model to Low Energies}

The results in Sec. II pertains to annihilation of the antiproton at
high energies with straight-line trajectories.  To extend the range of
the application to low energies, it is necessary to forgo the
assumption of straight-line trajectories.  We need to take into
account the modification of the trajectories due to residual Coulomb
and nuclear interactions that are additional to those between the
incident antiproton and an annihilated target nucleon.

The residual Coulomb and nuclear interactions affect the annihilation
process in different ways.  The Coulomb interaction is long-range, and
the trajectory of the antiproton is attracted and pulled toward the
target nucleus before the antiproton makes a contact with the nucleus
[Fig.\ 2(a)].  It leads to a large enhancement of the annihilation
cross sections at low energies \cite{Kuz94,Bat01}.  The nuclear
interaction is short-range, and it becomes operative only after coming
into contact with the nucleus.  The interactions modify the antiproton
momentum as it travels in the nuclear interior in which $\bar
p$-nucleon annihilation takes place.  As the basic antiproton-nucleon
annihilation cross section $\sigma_{\rm ann}^{\bar p p}$ as given by
(\ref{17a}) is strongly momentum-dependent, we need to keep track of
the antiproton momentum along the antiproton trajectory.

We shall work in the $\bar p$-$A$ center-of-mass system and shall
measure the antiproton momentum in terms of the relative momentum $\bb
p_{\bar p A}$ defined as
\begin{eqnarray}
\bb p_{\bar p A} = \frac{m_A \bb p_{\bar p} - m_{\bar p} \bb p_A}{m_{\bar p}+m_A},
\end{eqnarray}
where in the center-of-mass system with $\bb p_{\bar p}+\bb p_A=0$, we
have $\bb p_{\bar p}=\bb p_{\bar p A}$.

\subsection{Initial-State Coulomb Interaction} 
We consider the collision of an antiproton with a nucleus in the $\bar
p A$ center-of-mass system with a center-of-mass energy $E$.  The
initial antiproton momentum $\bb p_{\bar p A}$ is related to $E$ by
\begin{eqnarray}
E=\frac {\bb p_{{\bar p} A}^2}{2\mu}.
\end{eqnarray}
After the projectile antiproton travels along a Coulomb trajectory, it
makes contact with the nucleus at $r=R_c$ with a momentum $\bb p_{\bar
  p A}'$ determined by
\begin{eqnarray}
E&=& \frac{[\bb p_{{\bar p} A}']^2}{2\mu}+V_c(R_c),
\label{20}
\end{eqnarray}
where $V_c(R_c)$ is the Coulomb energy for the antiproton to be at
the nuclear contact radius $R_c$,
\begin{eqnarray}
V_c(R_c)=-\frac{(Z_A-1)\alpha}{R_c}, 
\label{28}
\end{eqnarray}
and $Z_A$ 
is the target charge number. 
 
The $\bar p p$ annihilation cross section arises from the nuclear and
Coulomb interaction between the antiproton and a target proton.  By
employing $\sigma_{\rm ann}^{\bar p p}$ as the basic element in the
multiple collision process in the extended Glauber model of
Eq.\ (\ref{5}), the parts of the Coulomb and nuclear interaction that
are responsible for the $\bar p p$ annihilation have already been
included.  Therefore in Eq.\ (\ref{20}) for $\bar p$-nucleus
annihilation, we are dealing with residual interactions that are
additional to those between the antiproton and the annihilated
nucleon.  Hence, Eq.\ (\ref{28}) for the residual Coulomb interaction
contains the coefficient ($Z_A-1$).

From angular momentum conservation, we have
\begin{eqnarray}
p_{{\bar p} A} b = p_{{\bar p} A}' b'. 
\end{eqnarray}
We obtain
\begin{eqnarray}
b=\frac{p_{{\bar p} A}'}{p_{{\bar p} A}} b'.
\label{31}
\end{eqnarray}
From Eq. (\ref{20}), we have 
\begin{eqnarray}
 p_{{\bar p} A}'
&=& p_{{\bar p} A}\sqrt{  1-\frac{V_c(R_c)}{E}},
\label{32}
\end{eqnarray}
which is the relative momentum of the antiproton in the $\bar p A$
system at the nuclear contact radius $R_c$.

 Starting now from the transverse coordinates $b'$ at nuclear contact
at $r=R_c$, one can follow the antiproton trajectory in the nuclear
interior.  This trajectory will be modified by residual interactions.
One can evaluate a thickness function $T'(b')$ by integrating the
nuclear density along the modified trajectory.  As the thickness
function $T'(b')$ is governed mainly the geometry of the target
nucleus, we therefore expect that $T'(b')$ will be characterized by a
length scale that will not be too different from the length scale in
the thickness function $T(b')$ without residual interactions.  To the
lowest order, it is reasonable to approximate $T'(b')$ by $T(b')$.
With such a simplifying assumption, the $\bar p A$ annihilation cross
section at an initial antiproton momentum $p_{\bar p A}$ is
\begin{eqnarray}
\Sigma_{\rm ann}^{\bar p A}(p_{\bar p A})  
&=&
\int d \bb b \biggl \{1 - \{1- T_{\bar p p}[\bb b'(\bb b)] \sigma_{\rm ann}^{\bar p p}(p_{\bar p A}')\}^Z
\nonumber \\
&&\times \{1- T_{\bar p n}[\bb b'(\bb b)] \sigma_{\rm ann}^{\bar p n}(p_{\bar p A}')\}^N \biggr \},~~~~
\end{eqnarray}
where we use a new symbol $\Sigma$ to indicate that this is the result
in an extended Glauber model for $\bar p$-nucleus annihilation,
modified to take into account the Coulomb interaction that changes the
antiproton momentum from initial $p_{\bar p A}$ to $p_{\bar p A}'$ at
the nuclear contact radius $R_c$.  We can carry out a change of
variable
\begin{eqnarray}
\Sigma_{\rm ann}^{\bar p A}( p_{\bar p A}) &\!=\!&
\!\int\!\! d \bb b' \frac{ b d b} {b' d b'} \biggl \{1 - \{1- T_{\bar p p}[\bb b'(\bb b)] \sigma_{\rm ann}^{\bar p p}(p_{\bar p A}')\}^Z
\nonumber \\
&&\times \{1- T_{\bar p n}[\bb b'(\bb b)] \sigma_{\rm ann}^{\bar p n}(p_{\bar p A}')\}^N \biggr \},~~~~
\label{41}
\end{eqnarray}
From Eqs. (\ref{31}) and (\ref{32}), the above Eq. (\ref{41}) becomes
\begin{eqnarray}
&&\Sigma_{\rm ann}^{\bar p A}( p_{\bar p A})
=
\frac{[p_{{\bar p} A}']^2}{p_{{\bar p} A}^2}
\int d \bb b'   \biggl \{1 - \{1- T_{\bar p p}[\bb b'(\bb b)] \sigma_{\rm ann}^{\bar p p}(p_{\bar p A}')\}^Z
\nonumber \\
&&~~~~~~~~~~~~~~~~\times \{1- T_{\bar p n}[\bb b'(\bb b)] \sigma_{\rm ann}^{\bar p n}(p_{\bar p A}')\}^N \biggr \}
\nonumber\\
&&~~=
\biggl \{  1-\frac{V_c(R_c)}{E}\biggr \}
\int d \bb b'   \biggl \{1 - \{1- T_{\bar p p}[\bb b'(\bb b)] \sigma_{\rm ann}^{\bar p p}(p_{\bar p A}')\}^Z
\nonumber \\
&&~~~~~~~~~~~~~~~~\times \{1- T_{\bar p n}[\bb b'(\bb b)] \sigma_{\rm ann}^{\bar p n}(p_{\bar p A}')\}^N \biggr \}
\label{27}
\end{eqnarray}
From the results in Eq.\ (\ref{6}), the above equation becomes
\begin{eqnarray}
\Sigma_{\rm ann}^{\bar p A}( p_{\bar p A})
&=&\biggl \{  1-\frac{V_c(R_c)}{E}\biggr \}
\sigma_{\rm ann}^{\bar p A}(\sigma_{\rm ann}^{\bar p p}(p_{\bar p A}')).
\label{36}
\end{eqnarray}

\begin{figure}[h]
 \includegraphics[width=0.4\textwidth]{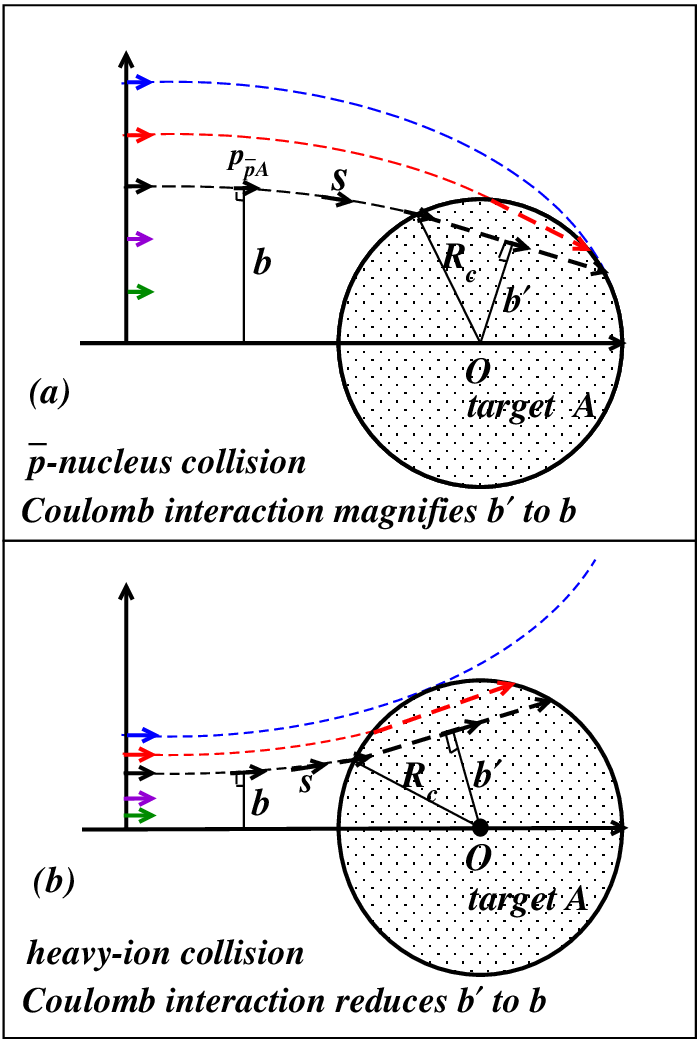}
\caption{ (Color online) Schematic picture of the bending of the
  antiproton trajectories $s$ as the antiproton approaches the
  nucleus: (a) Under the action of the attractive ${\bar p}$-nucleus
  Coulomb interaction, the impact parameter $b'$ in the nuclear
  interior is magnified to the initial impact parameter $b$ which
  determines the annihilation cross section.  (b) Under the action of
  a repulsive Coulomb interaction in heavy-ion collisions, the initial
  impact parameter $b'$ in the nuclear interior is reduced to the
  initial impact parameter $b$ which determines the reaction cross
  section. }
\end{figure}

With the above formulation, the initial-state Coulomb interaction can
be incorporated into the extended Glauber model as a mapping of the
initial impact parameter $b$ to the impact parameter $b'$ at nuclear
contact.  It can be pictorially depicted as a lens effect in Fig. 2.
The attractive Coulomb interaction between the antiproton and the
nucleus acts as a magnifying lens that magnifies the impact parameter
$b'$ at the contact radius $R_c$ to turn it into the initial impact
parameter $b$, with which the reaction or annihilation cross section
is measured [Fig.\ 2(a)].  The magnifying lens effect for the
attractive Coulomb interaction with $b > b'$ leads to a $\bar
p$-nucleus annihilation cross section greater than the geometrical
cross section for heavy nuclei in low-energy collisions, behaving as $
[1-V_c(R_c)/E] \sigma_{\rm ann}^{\bar p A}(\sigma_{\rm ann}^{\bar p
  p}(p_{\bar p A}'))$ as given in Eq.\ (\ref{36}), where $V_c(R_c)$,
the Coulomb energy at the nuclear contact radius $R_c$, is negative.

It is interesting to note in contrast that in heavy-ion collisions,
the repulsive Coulomb initial-state interaction acts as a reducing
lens that reduces the impact parameter $b'$ at contact to become the
initial impact parameter $b$ with $b < b'$ as illustrated in
Fig. 2(b).  The lens effect for a repulsive Coulomb interaction leads
to a reaction cross section reduced from the geometrical cross section
$\pi R_c^2$ to $[1-V_c(R_c)/E]\pi R_c^2$, where $V_c(R_c)$ is the
Coulomb energy at the Coulomb barrier and is positive
\cite{Bla52,Won73,Won12}.  Therefore, one obtains the unifying picture
that for both the $\bar p$-nucleus and heavy-ion collisions, the
initial-state Coulomb interaction act as a lens, leading to the same
Coulomb modifying factor $[1-V_c(R_c)/E]$.

\subsection{Change of the Antiproton Momentum in the Nucleus Interior}

Because the basic $\bar p p$ annihilation cross section is strongly
momentum dependent, there is, however, an additional important amendment
we need to make.  In the presence of the nuclear and Coulomb
interactions in the nuclear interior, the antiproton momentum $p_{\bar
  p A}'$ at the contact radius is changed to the momentum $p_{\bar p
  A}''$ in the interior of the nucleus.  The $\bar p p$ annihilation
occurs inside the nucleus at a momentum $p_{{\bar p} A}''$.  It is
necessary to modify $\Sigma_{\rm ann}^{\bar p p}$ in Eq.\ (\ref{36})
to take into account this change of the antiproton momentum by
replacing $p_{\bar p A}'$ with $p_{\bar p A}''$.

The antiproton momentum $p_{{\bar p} A}''$ is a function of the radial
position $r$ and is related to $p_{\bar p A}$ by the energy condition:
\begin{eqnarray}
E=\frac{\bb p_{{\bar p} A}^2}{2\mu}=\frac{(\bb p_{{\bar p} A}'')^2}{2\mu}+V_c(r)
+ V_n(r).
\end{eqnarray}

In order to obtain an analytical formula for general purposes for our
present approximate treatment, it suffices to consider average
quantities and use the root-mean-square average $\langle (\bb p_{{\bar
    p} A}'')^2\rangle^{1/2}$ given by
\begin{eqnarray}
E
=\frac{\langle (\bb p_{{\bar p} A}'')^2\rangle }{2\mu}+\langle V_c(r)\rangle 
+ \langle  V_n(r)\rangle ,
\end{eqnarray}
where $\langle V_c(r)\rangle$ and $\langle V_n(r)\rangle$ are the
average interior Coulomb and nuclear interactions, respectively.  From
the above equation, we can then approximate $p_{{\bar p} A}''$ by the
average root-mean-squared momentum $\langle (\bb p_{{\bar p}
  A}'')^2\rangle^{1/2}$ in the interior of the nucleus,
\begin{eqnarray}
 p_{{\bar p} A}''
&\sim & p_{{\bar p} A}\sqrt{1-\frac{\langle V_c(r)\rangle 
+ \langle  V_n(r)\rangle}{ E}}.
\label{38}
\end{eqnarray}

The $\bar p$-nucleus annihilation cross section $ \Sigma_{\rm
  ann}^{\bar p A}$ is therefore modified by changing $p_{\bar p A}'$in
Eq.\ (\ref{36}) to $p_{\bar p A}''$ to become
\begin{eqnarray}
\Sigma_{\rm ann}^{\bar p A}( p_{\bar p A})
&=&\biggl \{  1-\frac{V_c(R_c)}{E}\biggr \}
\sigma_{\rm ann}^{\bar p A}(\sigma_{\rm ann}^{\bar p p}(p_{\bar p A}'')).
\label{39}
\end{eqnarray}

The experimental data of $\sigma_{\rm ann}^{\bar p p}$ and
$\Sigma_{\rm ann}^{\bar p A}$ are presented as a function of $p_{{\bar
    p}{\rm lab}}$ for a fixed proton or nucleus target at rest.
Accordingly, we convert the antiproton momenta in the center-of-mass
system in th above Eq.\ (\ref{39}) to those in the laboratory system with
fixed targets as
\begin{eqnarray}
 p_{\bar p A}&=&\frac{A}{A+1}  p_{{\bar p}{\rm lab}},\\
p_{{\bar p} A}'' &=&\frac{A}{A+1} p_{{\bar p}{\rm lab}}'' ,
\end{eqnarray} 
and Eq. (\ref{38}) gives
 \begin{eqnarray}
p_{\bar p {\rm lab}}''
&=&  p_{\bar p \rm lab}\sqrt{1-\frac{\langle V_c(r)\rangle 
+ \langle  V_n(r)\rangle}{ E}}.
\label{42}
\end{eqnarray}
Therefore, the antiproton-nucleus annihilation cross section
$\Sigma_{\rm in}^{\bar p A}( p_{\bar p {\rm lab}})$ in the extended
Glauber model for an antiproton with an initial momentum $p_{\bar p {\rm
    lab}}$ is given in the following compact form
\begin{eqnarray}
\Sigma_{\rm ann}^{\bar p A}( p_{\bar p {\rm lab}}) &=& \biggl
\{1-\frac{V_c(R_c)}{E}\biggr \} \sigma_{\rm in}^{\bar p
  A}(\sigma_{\rm ann}^{p\bar p}(p_{\bar p {\rm lab}}'')),
\label{43}
\end{eqnarray}
where $\sigma_{\rm in}^{\bar p A}(\sigma_{p\bar p}(p_{\bar p {\rm
    lab}}''))$ is given by Eq.\ (\ref{9}) for light nuclei with a
Gaussian thickness distribution, and by Eq.\ (\ref{19}) for heavy
nuclei with a sharp-cut-off thickness function, with the basic
quantity $\sigma_{\rm ann}^{\bar p p}$ in Eqs.\ (\ref{10}) or
(\ref{20a}) evaluated at $p_{\bar p {\rm lab}}''$ given in terms of
$p_{\bar p {\rm lab}}$ by Eq. (\ref{42}).  The Coulomb energy at
contact $V_c(R_c)$ is given by Eq.\ (\ref{28}), and the average
$\langle V_c(r) \rangle$ in the interior of the nucleus $(Z_A,A)$ with
a radius $R_A$ in Eq.\ (\ref{42}) is given by
\begin{eqnarray}
\langle V_c(r)\rangle & = & -3
(Z_A-1)\alpha\left(\frac{5R^2_c-R^2_A}{10R^3_c}\right).
\label{44}
\end{eqnarray} 

In our application of the extended Glauber model, concepts such as the
contact radius $R_c$ and $\langle V_c(r)\rangle $ are simplest for a
uniform density distribution.  For the evaluation of these quantities
in the case of small nuclei with a Gaussian thickness function, we
shall approximate the Gaussian as a uniform distribution [only for the
  purpose of calculating $R_c$ and $\langle V_c(r)\rangle $ in
  Eqs.\ ({\ref{28}), (\ref{42}), and (\ref{44})] with an equivalence
  between $R_c$ and $\beta$.  We note that the dimensionless quantity
  $f$ in Eq.\ (\ref{10}) for a Gaussian thickness function and the
  quantity $F$ in Eq.\ (\ref{20a}) for the sharp-cut-off distribution
  have the same physical meaning.  It is reasonable to equate the
  corresponding quantities $2\pi\beta^2$ in Eq.\ (\ref{10}) with the
  corresponding quantity $2\pi R_c^2/3$ in Eq.\ (\ref{20a}), leading
  to the approximate equivalence
\begin{eqnarray}
R_c({\rm for~Gaussian~distribution})\sim \sqrt{3} \beta,
\end{eqnarray}
and similarly, 
\begin{eqnarray}
R_A({\rm for~Gaussian~distribution})\sim \sqrt{3} \beta_A.
\end{eqnarray}
These equivalence relations enable us to obtain the Coulomb factor
$[1-V_c(R_c)/E]$ in (\ref{43}) and the antiproton momentum change from
$p_{\bar p {\rm lab}}$ to $p_{\bar p {\rm lab}}''$ in Eq.\ (\ref{42}),
for light nuclei with Gaussian thickness functions.

\section{Comparison of the Extended Glauber Model with Experiment} 

The central results of the extended Glauber model consist of
Eq.\ (\ref{39}) or (\ref{43}) and their associated supplementary
equations.  With the basic $\bar p p$ annihilation cross section
$\sigma_{\rm ann}^{\bar p p}$ and its momentum dependence well
represented from experimental data by Eq.\ (\ref{17a}), as discussed in
Sec. III, it is only necessary to specify the residual nuclear
interaction $\langle V_N\rangle$ and the nuclear geometrical
parameters in Eq.\ (\ref{39}) to obtain the $\bar p$-nucleus
annihilation cross section.  For light nuclei (with $A < 40$), we use
a Gaussian density distribution with the Gaussian thickness function
in Eq.\ (\ref{6}), with geometrical parameters $\beta_A=r_0'
A^{1/3}/\sqrt{3}$ and $\beta_B=0.68$.  We find that $ 1.10 < r_0'  <1.20$  fm
gives a good description.  For the heavy nuclei (with $A >$ 40), we
use a uniform density distribution with the sharp-cut-off thickness
function in Eq.\ (\ref{17}) with geometrical parameters $R_A=r_0
A^{1/3}$ and $R_B$ = 0.8 fm The radius parameter $r_0=1.04$ fm fit the
data well.

In Fig.\ 1, the solid curve for the proton target nucleus is from the
$\sigma_0/v$ phenomenological representation of the $\bar p p$
annihilation cross section in Eq.\ (\ref{17a}).  The other curves are
the results from the extended Glauber model using the basic $\bar p p$
annihilation cross section as input data.  The solid curves are for
Gaussian density distributions and the dashed curves are for uniform
density distributions.  The fitting parameters that give the
theoretical $\bar p A$ annihilation cross sections in Fig. 1 are
listed in Table I.

\begin{table}  [h]
\caption{Fitting parameters}
\begin{tabular}{cccc}
\hline
Nuclei & Gaussian &  Uniform &  ~$\langle V_n\rangle $(MeV) \\
          &~~ $r^\prime_0$(fm)~~ &  ~~$r_0$(fm) ~~&  \\
\hline
    $^2$H          &   1.20  &            &   -1.0 \\
    $^4$He &  1.20   &           &   -4.0 \\
    Be       &     1.20   &          &  -10.0 \\
    C          &   1.20  &           &  -15.0  \\
    Ne        &   1.10  &           &  -25.0  \\
    Al         &  1.10   &           &  -30.0  \\
    Ni         &           & 1.04    &  -30.0    \\
    Cu        &          &  1.04    &  -30.0  \\
    Cd        &           &  1.04    & -35.0  \\
    Sn        &           &  1.04    & -35.0  \\
    Pt         &           &  1.04    & -35.0   \\
    Pb        &           &  1.04    &  -35.0  \\
\hline
\end{tabular}
\end{table}

\begin{figure}[h]
  \centering
    \includegraphics[scale=0.4]{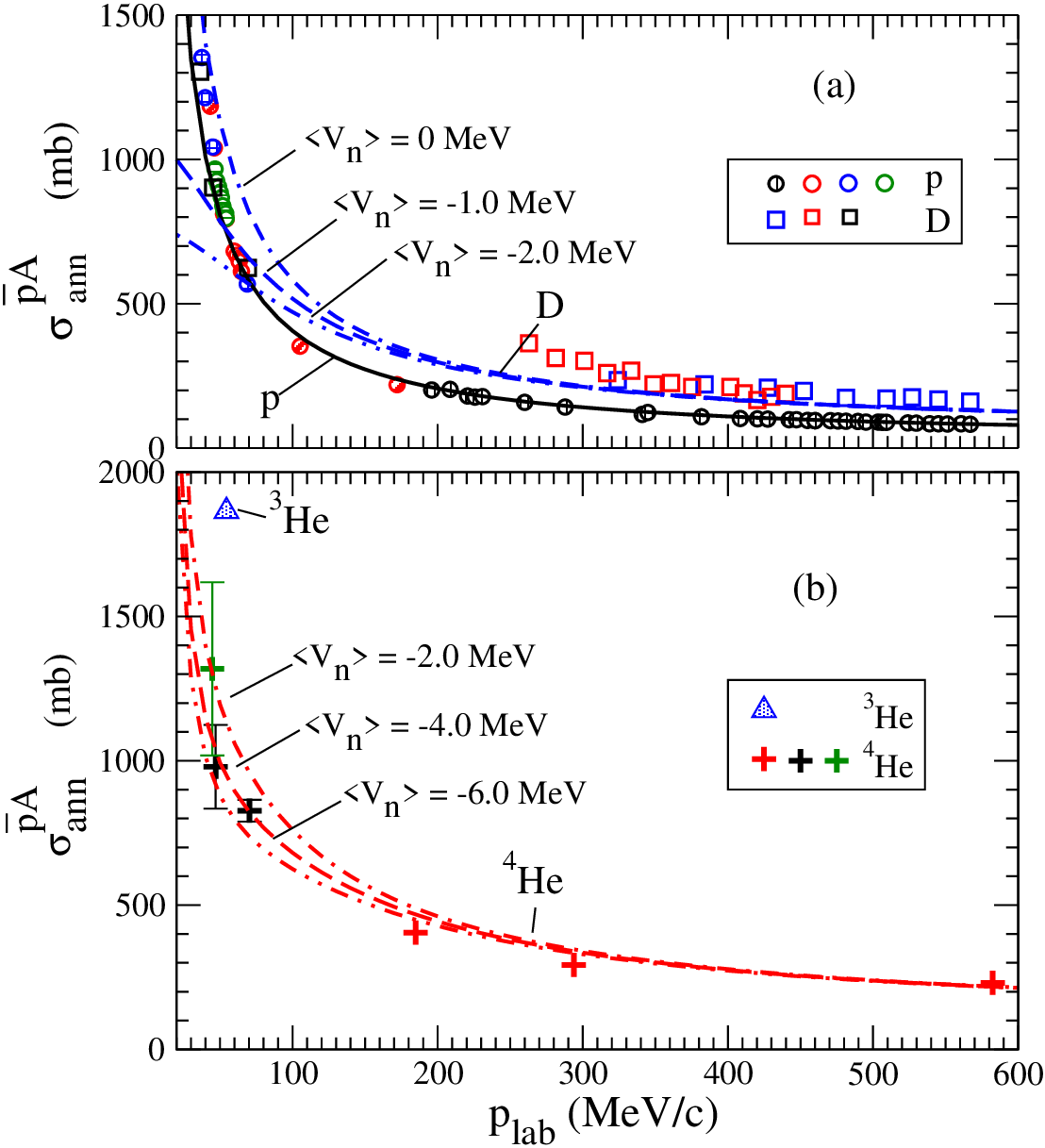}
\caption{ (Color online) $\bar{p} $ annihilation cross sections of (a)
  $p$ and D nuclei and (b) $^4$He nucleus. The experimental data are
  from Ref. \cite{Bia11}. The $\bar p p$ curve is from the
  $\sigma_0/v$ representation of Eq.\ (\ref{17a}), and the other curves
  are from the extended Glauber model. }
\end{figure}

The comparison of the extended Glauber model with the experimental
data in Fig.\ 1 indicates that although the fits are not perfect, the
extended Glauber model captures the main features of the annihilation
cross sections for all energies and mass numbers.  We will mention a
few of the notable features of the data and the corresponding
explanations in the extended Glauber model.

We examine first $\bar pA$ annihilation at high energies.  In these
high-energy annihilations, the momentum dependence of the basic
$\sigma_{\rm ann}^{\bar p p}$ is not sensitive to the antiproton
momentum change arising from the residual nuclear and Coulomb
interactions $\langle V_c \rangle$ and $\langle V_N \rangle$.  The
initial-state Coulomb interaction energy $V_{c}(R_c)$ is also small in
comparison with the incident energy $E$.  As a consequence, the
corrections due to the Coulomb and nuclear interactions are small for
high-energy collisions.  At these high energies, the antiproton makes
multiple collisions with target nucleons and probes the granular
property of the nucleus when the spacing between the nucleons is large
compared with the dimension of the $\bar p$ probe, similar to the
additive quark model in meson-meson collisions \cite{Lev65,Won96}.
The experimental $\Sigma_{\rm ann}^{\bar p ^2H}/\sigma_{\rm ann}^{\bar p
  p}$ is consistently slightly greater than the theoretical prediction
of $\Sigma_{\rm ann}^{\bar p ^2H}/\sigma_{\rm ann}^{\bar p p}$$\sim$ 2.
This indicates that while the Glauber model for the antiproton
annihilation of the deuteron may be approximately valid, future
investigations will need to include additional refinements to get a
better description for the antiproton-deuteron annihilation.  There
are not many high-momentum data points for ${}^4$He, and the only
high-momentum data point at 600 MeV/$c$ gives reasonable agreement with the
Glauber model results.

As the number of nucleons increases in high-energy annihilations, the
Glauber model results of Eqs.\ (\ref{9})
[or  (\ref{21a}) for $\sigma_{\rm ann}^{\bar p n}$$\ne$$\sigma_{\rm ann}^{\bar p p}$] 
and   (\ref{19})
[or (\ref{22a}) for $\sigma_{\rm ann}^{\bar p n}$$\ne$$\sigma_{\rm ann}^{\bar p p}$]  
give a $\bar p A$ annihilation cross section proportional to $\beta^2$ for a Gaussian
thickness distribution and to $R_c^2$ for a uniform density
distribution.  Both $\beta^2$ and $R_c^2$ vary as $A^{2/3}$.  Hence,
the $\bar p A$ annihilation cross approaches the black-disk limit of
$A^{2/3}$ limit as the nuclear mass number increases. The comparison
in the high-energy region in Fig.\ 1 shows that the experimental data
agree with predictions for nuclei across the periodic table,
indicating that the experimental data indeed reach this black-disk
limit of $A^{2/3}$ in the heavy-nuclei region, However, there are
discrepancies for Pb at 700 MeV/$c$, which may need need to be re-checked
experimentally, as the other data points for large nuclei appear to agree with 
theoretical predictions.

\begin{figure}[h]
  \centering
    \includegraphics[scale=0.4]{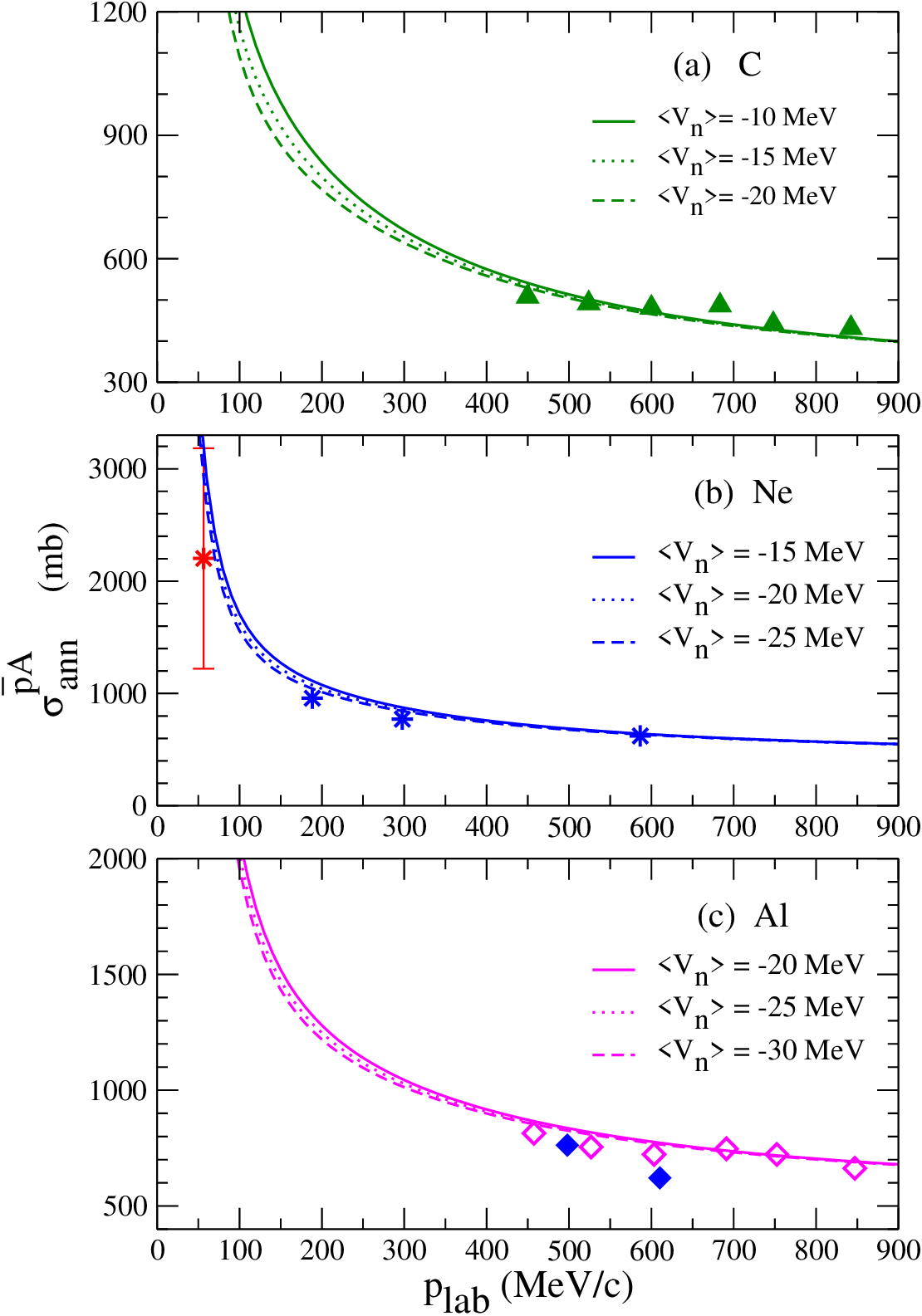}
\caption{ (Color online) $\bar{p} $ annihilation cross sections of (a)
  C nucleus, (b) Ne nucleus, and (c) Al nucleus.  Solid curves give
  results from the extended Glauber model and the data points are from
  the compilation of Ref. \cite{Bia11}.}
\end{figure}

The situation at the low-energy region is more complicated.  In
addition to the multiple collision process, the Coulomb and nuclear
interactions also come into play.  We can examine the data for $p$, $^2$H,
and He more closely in Fig.\ 3 in a linear plot in both the low and
high energy regions.  As shown in Figs.\ 1 and 3(a), the basic cross
section $\sigma_{\rm ann}^{\bar p p}$ is of order 1000 mb at the low
momentum of $p_{\bar p {\rm lab}}$$\sim$20 MeV/$c$, corresponding to a
effective annihilation radius between $\bar p$ and $p$ of $R_{\bar p
  p}=(\sigma_{\rm ann}^{\bar p p}/\pi)^{1/2}$=6.8 fm.  The large cross
section and annihilation radius arise from the magnifying lens effect
of the attractive initial-state Coulomb interaction that magnifies the
proton radius as seen by the incoming antiproton, as discussed in
Sec. III.  In this case, as $R_{\bar p p} \gg R_{\rm deuteron} {\rm
  ~or~} R_{\rm antiproton}$, the quantity $f$ in Eq.\ (\ref{9}) is
close to unity.  According to the analysis given in Eq.\ (14) of
Sec. III, the Glauber model with $f\sim 1$ would predict a ratio of
$\Sigma_{\rm ann}^{\bar p ^2H}/\sigma_{\rm ann}^{\bar p p}\sim 3/2$ for
low-energy annihilations in the multiple collision process.

\begin{figure}[h]
  \centering
    \includegraphics[scale=0.28]{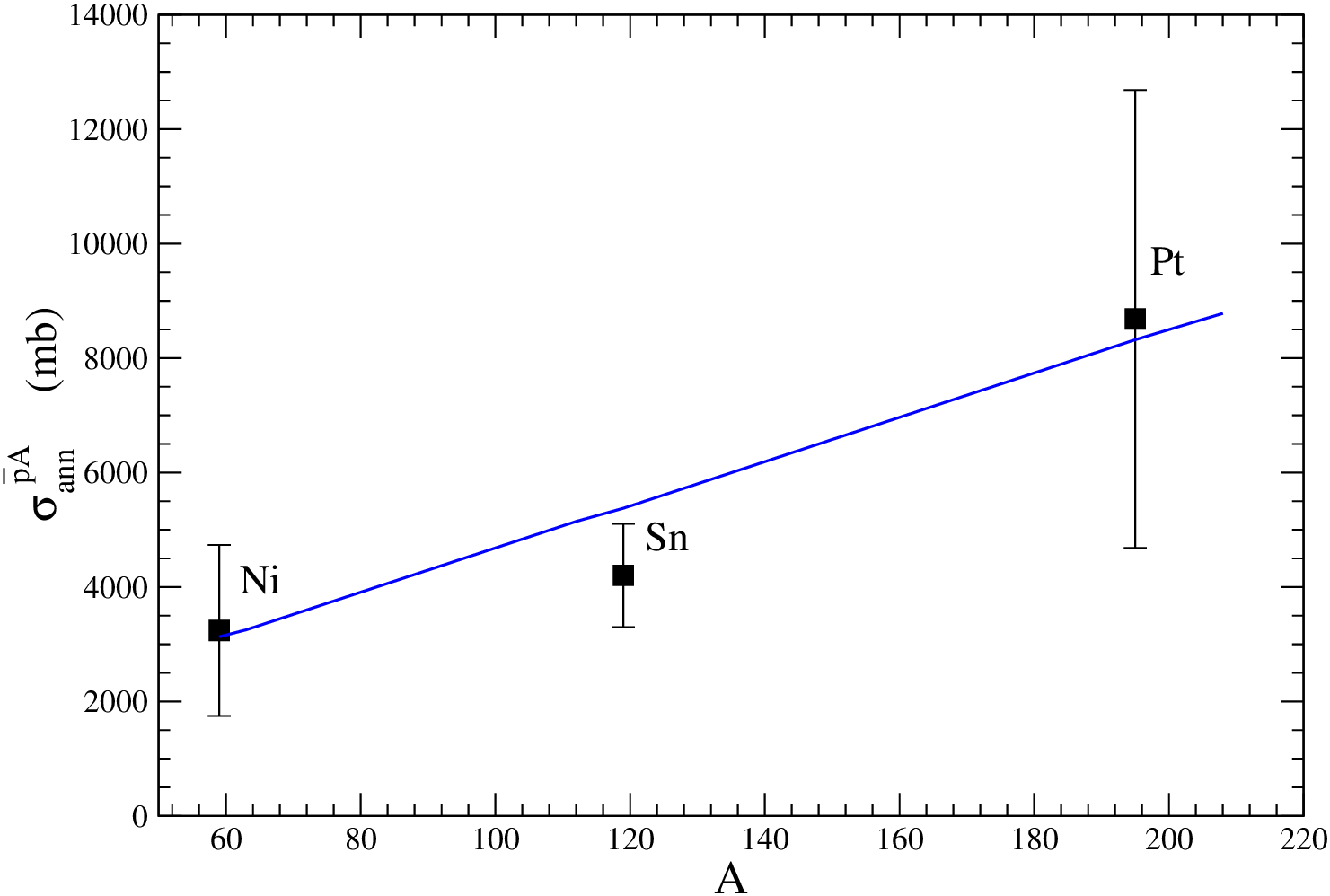}
  \caption{ (Color online) $\bar p $ annihilation cross sections on Ni, Sn and Pt
    nuclei at $p_{\bar p {\rm lab}}$ = 100 MeV/$c$.  The solid curve
    gives results from the extended Glauber model and the data points
    are from Ref. \cite{Bia11}.}
\end{figure}

Experimentally, for low-energy annihilations $\Sigma_{\rm ann}^{\bar p
 {}^2H}/\sigma_{\rm ann}^{\bar p p}$ is of order unity.  In the extended
Glauber model, the reduction of the ratio $\Sigma_{\rm ann}^{\bar p
 {}^2H}/\sigma_{\rm ann}^{\bar p p}$ arises from the combination of two
effects: (i) the increase of the antiproton momentum inside the
nucleus due to the attractive residual interactions and (ii) the
momentum dependence of the basic $\sigma_{\rm ann}^{\bar p p}$
decreases sensitively as a function of an increase in antiproton
momentum.  The combined effects bring the ratio $\Sigma_{\rm
  ann}^{\bar p{}^2H}/\sigma_{\rm ann}^{\bar p p}$ from 3/2 to about
unity.  In Fig.\ 3, we show the variation of the $\bar p {}^2$H and $\bar
p {}^4$He cross sections as a function of $p_{\bar p {\rm lab}}$ for
different residual nuclear interactions $\langle V_N\rangle$.  There
is a great sensitivity of the $\bar p$-annihilation cross section on
$V_N$ in the low energy region for the lightest nuclei for which the
Coulomb interaction is weak.  However, as the target charge number
increases, the Coulomb interaction becomes stronger and the $\bar
p$-annihilation cross section becomes less sensitive to the strength
of the nuclear interaction $\langle V_N\rangle$, as shown in Fig.\ 4
for $\bar p$C, $\bar p$Ne, and $\bar p$Al collisions.

As the target charge number $Z$ increases further in the low-energy
region, the cross section increases substantially.  For example, the
cross section reaches a value of about 9000 mb for the Pt nucleus,
corresponding to an annihilation radius of $R_{\rm
  ann}=(\sigma_{\rm ann}^{\bar p A}/\pi)^{1/2}$=17 fm.  Again, such a
large annihilation radius arises from the Coulomb magnifying lens
effect that magnifies the nuclear radius of Pt as seen by the incoming
antiproton.  The attractive Coulomb
interaction as well as the nuclear interaction also changes the
momentum of the antiproton inside the nucleus.  The combined effect of
the Coulomb initial-state interaction, the change of the antiproton
momentum inside a nucleus, together with the Glauber multiple
collision process of individual antiproton-nucleon annihilation, give
a good description of the cross sections for the heaviest nuclei at
low energies, as shown in Figs.\ 1 and 5.

There are only a few cases where the experimental data points deviate
from the general trend and the theoretical predictions. In particular,
there are discrepancies for $\bar p{}^2$H at $p_{\rm lab}$$\sim$260-600
MeV/$c$, $\bar p$(${}^3$He) at 55 MeV/$c$, $\bar p $Pb at 700 MeV/$c$, and
$\bar p$Be at high momenta.  It will be necessary to examine the
origins for the discrepancies by theoretical refinements or
experimental remeasurements for these cases in the future.

\section{Summary and Discussions}

We have extended the high-energy Glauber model to low-energy
annihilation processes after taking into account the effects of Coulomb
and nuclear interactions, and the change of the antiproton momentum
inside a nucleus. The result is a compact equation (\ref{43}) [or the
equivalent (\ref{39})] with supplementary equations that capture the
main features of the annihilation process and provide a simple
analytical way to analyze antiproton-nucleus annihilation cross
sections.

We can properly respond to the specific questions we raised in the
Introduction.  With respect to the mass dependence at high antiproton
incident momenta, the cross sections increase almost linearly with the
mass number $A$ for the lightest nuclei, but with approximately
$A^{2/3}$ as the mass number increases because the basic process is a
Glauber multiple collision process of the antiproton passing through a
target of individual nucleons.  In the low antiproton momentum region,
the annihilation cross sections do not rise with $A$ as anticipated
but are comparable for $p$, $^2$H, and He, because the residual nuclear
interaction causes an increase in the antiproton momentum inside the
nucleus and the increases in antiproton momentum leads to a decrease
in the basic $\bar p$-nucleon annihilation cross section.  As the
charge number $Z$ increases, the initial-state Coulomb interaction
magnifies the target nucleus and the $\bar p A$ annihilation cross
section at low energies become subsequently greatly enhanced in the
heavy nuclei region.  With respect to the energy dependence, the
energy dependence of the $\bar p A$ annihilation cross sections is
intimately related to the energy-dependence of the $\bar p p$ cross
sections.

The simple picture we have presented can be refined, and individual
nuclear properties can be revealed, with new data that fill in the
gaps in Fig. 1.  The deviations of experimental data with theoretical
predictions for a few cases also call for a reexamination of both the
experimental measurements as well as theoretical refinements.

The extended Glauber model may find future applications in similar
problems such as in the collision of mesons or baryons with nuclei 
at high energies as well as low energies.

\vspace*{0.5cm}
\centerline {\bf Acknowledgment}

\vspace*{0.2cm} This research was supported in part by the Division of
Nuclear Physics, U.S. Department of Energy.

\end{document}